\documentstyle[12pt,a4,epsfig]{article}
             \let\section=\subsection 
	     \let\subsection=\subsubsection                
               
\begin{document}

\begin{center}
  {\large \bf STABILITY OF THE HEAVIEST NUCLEI}\\[2mm]
  {\large \bf ON SPONTANEOUS--FISSION}\\[2mm]
  {\large \bf AND ALPHA--DECAY$^{\textstyle \bf \dagger \, \ddagger}$}\\[5mm]

  {\large A.~Staszczak, Z.~\L{}ojewski, A.~Baran, B.~Nerlo--Pomorska\\
  and K.~Pomorski}\\[5mm]

  {\small \it  Department of Theoretical Physics, M. Curie--Sk\l{}odowska
  University \\
  pl. M. Curie--Sk\l{}odowskiej 1, 20--031 Lublin, Poland\\[8mm] }
\end{center}

{
 \renewcommand{\thefootnote}
 {\fnsymbol{footnote}}
 \footnotetext[2]{
  This work was supported in part by Polish State Committee for Scientific 
  Research (KBN) under Contract No. 2 P03B 049 09}
}

{
 \renewcommand{\thefootnote}
 {\fnsymbol{footnote}}
 \footnotetext[3]{
  Talk presented at Third International Conference on Dynamical Aspects of
  Nuclear Fission DANF`96 in \v{C}ast\'{a}--Papierni\v{c}ka,
  Slovak Republic, Aug 30--Sep 4, 1996}
}

\begin{abstract}\noindent
Spontaneous--fission half--lives $(T^{SF}_{1/2})$ of the heaviest even--even
nuclei are evaluated and compared with their alpha--decay mode.
Calculations of $T^{SF}_{1/2}$ are performed in the dynamical way
with potential energy obtained by the macroscopic--microscopic method 
and the inertia tensor obtained by the cranking approximation.
The alpha--decay half--lives $(T^{\alpha}_{1/2})$ are calculated for 
the same region of nuclei by use of the Viola and Seaborg formula.
The ground--state properties such as mean square radii and electric 
quadrupole moments are also studied.

From the analysis of $T^{SF}_{1/2}$ it is found that a peninsula
of deformed metastable superheavy nuclei near $^{268}_{106}{\rm Sg}_{162}$
is separated from an island of the spherical superheavy, around the doubly
magic nucleus $^{298}114_{184}$, by a trench in the vicinity of neutron
number N=170.
\end{abstract}


\section{INTRODUCTION}
Much progress has been made recently in the synthesis of very heavy nuclides.
Deformed superheavy nuclides with proton numbers Z=108, 110, 111 and 112
have been discovered through reactions of cold--fusion at GSI
\cite{Hof1,Hof2,Hof3} and hot--fusion in Dubna \cite{Laz1,Laz2}. In 1995, 
first chemical separations of element 106 were performed \cite{T}.
After these successes, the accurate calculations of the lifetimes of 
nuclei situated in the upper--end of the isotopic chart became a new 
challenge of nuclear theory.

The objective of the present paper is to study spontaneous--fission
$(T^{SF}_{1/2})$ and $\alpha$-decay $(T^{\alpha}_{1/2})$ half--lives for 
even-even nuclei with proton number Z=100--114 and neutron number N=142--180.
This relatively broad region of nuclei contains experimentally well known
nuclides with Z$\leq$104, the deformed superheavy with Z$\geq$106 and
transitional nuclei close to a hypothetical island of spherical
superheavy elements situated around the nucleus $^{298}114_{184}$.
For all these nuclei the action integrals 
describing probability of SF are minimized using multi--dimensional
dynamic--programming (MDP) method based on WKB approximation within the same
deformation space.

Much attention is paid to find the optimal deformation space. We examine
a~relatively reach collection of nuclear shape parameters 
($\beta_{\lambda}$, with $\lambda$=2,3,4,5,6 and 8), as well as the pairing 
degrees of freedom (i.e. proton $\Delta_{p}$ 
and neutron $\Delta_{n}$ pairing gaps). The optimal collective space
$\{\beta_{2}, \beta_{4}, \beta_{6}, \Delta_{p}, \Delta_{n}\}$ 
is found by comparison of the calculated $T^{SF}_{1/2}$ of Fm isotopes
with their experimental values.

Alpha--decay is one of the most predominant modes of decay
of superheavy nuclei. All recently discovered superheavy elements with atomic
numbers Z$\geq$107 were identified from their $\alpha$-decay chains.
Calculations of $T^{\alpha}_{1/2}$ are easier to perform than $T^{SF}_{1/2}$.
The half--life for $\alpha$-decay depends primarily upon the release energy 
$Q_{\alpha}$, which is given only by appropriate difference of ground--state 
masses.
To better characterized properties of the nuclei in investigated
region, we also calculated their ground--state electric quadrupole moments
and mean square radii.
A part of the results of the analysis has been presented earlier
\cite{LS1,SL1,LS2}.

The description of the method and details of the calculations 
are given in Sec.~2, the results and discussion are presented in Sec.~3.

\section{DESCRIPTION OF THE METHOD}

\subsection{Collective variables}
Let us consider a set of $\lambda$ collective variables 
$(X_1,X_2,...,X_{\lambda})\equiv X$, 
then the classical collective Hamiltonian can be written

\begin{equation}
H= \frac{1}{2}\sum_{k,l}^{\lambda} B_{kl}(X)\dot{X}_{k}\dot{X}_{l}+V(X)\,,
\end{equation}

\noindent
where $V(X)$ is the collective potential energy. The tensor of effective mass
(inertia tensor) $B_{kl}$ is symmetric and all its components may depend on
collective variables. The choice of the collective variables is arbitrary
but caution must be exercised.

Starting from the single--particle motion of A=Z+N nucleons, a state of 
nucleus is defined by an average potential fixed by a set of parameters. 
These parameters are good candidates for collective variables.
In the presented paper we used the single--particle Hamiltonian ($H_{s.p.}$)
consisted of a deformed Woods--Saxon Hamiltonian 
and a residual pairing interaction treated in the BCS approximation.
In our model we discussed only axially symmetric
deformations of $H_{s.p.}$, i.e. a nuclear radius was expanded in terms of
spherical harmonics $Y_{\lambda 0}(cos\vartheta)$

\begin{equation}
R(\vartheta) = R_{0}(\beta_{\lambda})\,[1 + \sum_{\lambda=2}
^{\lambda_{max}} \beta_{\lambda} Y_{\lambda 0}(cos\vartheta)]\,,
\end{equation}

\noindent
where $\beta_{\lambda}$ is the set of deformation parameters up to 
$\lambda_{max}$ multipolarity and the dependence of $R_{0}$ on
$\beta_{\lambda}$
is determined by the volume--conservation condition.

Besides the deformation parameters, the other candidates for collective 
variables are parameters connected with the pairing interaction.
In the usual BCS approximation we have two such parameters:
the pairing Fermi energy ($\lambda$) and the pairing energy--gap~($\Delta$).

In the presented paper we chose the proton $\Delta_{p}$ and neutron
$\Delta_{n}$ pairing gaps as the additional collective variables.
The significant role of these so--called pairing vibrations on 
penetration of the fission barrier has been first discussed by
Moretto {\em et al.} in Ref. \cite{MB} (see, also \cite{SB,SPP,PPS,D}).

Finally, the set of collective variables consists of the nuclear shape
parameters ($\beta_{\lambda}$) and the pairing degrees of freedom 
($\Delta_{p}, \Delta_{n}$). These variables spann the multi--dimensional 
deformation space $\{X_{\lambda}\}$ within which we shall describe a fission
process.

\subsection{Inertia tensor}
The inertia tensor $B_{kl}$ describes the inertia of a nucleus with 
respect to changes of its {\em shape}. It also plays a role similar to the
metric tensor in deformation space $\{X_{\lambda}\}$.
Its components for multipole vibrations as well as pairing vibrations
can be evaluated in the first order perturbation approximation 
\cite{Belya}

\begin{equation}
B_{kl}(X) = 2\hbar^2 \sum_{m}
  \frac{\langle 0|{\partial}/{\partial X_k}|m\rangle
        \langle m|{\partial}/{\partial X_l}|0\rangle}
	{{\cal E}_m - {\cal E}_0}\,, 
\end{equation}

\noindent
where $|0\rangle$ and $|m\rangle$ denote a ground state and
the excited states of the nucleus with the corresponding energies
${\cal E}_0$ and ${\cal E}_m$.
For even--even nuclei the excited states can be identified with the two 
quasi-particle excitations ${\cal E}_m=E_\mu+E_\nu$.
After transformation to the quasi-particle representation
the corresponding formula takes the following compact form \cite{GP}

\begin{equation}
B_{kl}(X) = 2\hbar^2 \sum_{\mu,\nu} P^{k \, \ast}_{\mu\nu}(X)
(E_\nu + E_\mu)^{-1} P^l_{\mu\nu}(X)\,,
\end{equation}
where for the shape deformations

\begin{equation}
   P^k_{\mu\nu}(\beta)
= -\frac{\langle \mu|\frac{\partial{H_{s.p.}}}{\partial\beta_k}|\nu\rangle}
{E_\mu+E\nu}
(u_\mu v_\nu + u_\nu v_\mu)-\frac{1}{2}
\delta_{\mu\nu} (\frac{\Delta}{E_\mu^2}\frac{\partial\lambda}
{\partial\beta_k}
 + \frac{e_\mu-\lambda}{E_\mu^2}\frac{\partial\Delta}{\partial\beta_k})
\end{equation}
and in the case of pairing degrees of freedom

\begin{equation}
   P^k_{\mu\nu}(\Delta)=
   \delta_{\mu\nu} \frac{(e_\mu-\lambda)+\Delta
   \frac{\partial\lambda}{\partial\Delta}}{2E_\mu^2}\,.
\end{equation}

\noindent
Here $v_\mu$, $u_\mu$ are the pairing occupation probability
factors, $e_\mu$ are the single--particle energies of $H_{s.p.}$
and the $E_\mu=[(e_\mu - \lambda)^2 + \Delta^2]^{1/2}$
is the quasi--particle energy corresponding to $|\mu\rangle$ 
state. The above expression is equivalent to a commonly used 
formula developed by Sobiczewski {\em et al.} in Ref. \cite{Sob69}.

The components of inertia tensor are strongly affected by single--particle
and pairing effects. The relation between the energy--gap parameter
$\Delta$, the effective level density at the Fermi energy
$g_{eff}(\lambda)$ and the diagonal components of inertia tensor 
$B_{kk}$ can be showed
in terms of the uniform model \cite{Funny}

\begin{equation}
B_{kk} \sim const.\frac{g_{eff}(\lambda)}{\Delta^2}
| \langle \partial H_{s.p.}/ \partial \beta_k \rangle |^2\,. 
\label{delb}
\end{equation}

\noindent
This strong dependence of the inertia tensor on the pairing energy--gap
allows to expect considerable reduction of the spontaneous--fission
half--life values.

\subsection{Collective energy}
The collective energy $V$ is calculated for a given nucleus by the
macroscopic--microscopic model developed by Strutinsky \cite{Strut}:

\begin{equation}
V=E_{\rm macr}(\beta)+\delta{}E_{\rm shell}(\beta)
+\delta{}E_{\rm pair}(\beta,\Delta)\,.
\end{equation}

\noindent
For the macroscopic part $E_{\rm macr}$ we used the
Yukawa--plus--exponential model \cite{KN}. 
The so--called microscopic part, consisted of the shell
$\delta{}E_{\rm shell}$ and pairing $\delta{}E_{\rm pair}$
corrections, was calculated on the basis of single--particle spectra of
Woods--Saxon Hamiltonian \cite{CDN}.

The one--body Woods--Saxon Hamiltonian consists of the kinetic energy term
$T$, the potential energy $V^{WS}$,
the spin-orbit term $V^{WS}_{so}$ and the Coulomb potential $V_{Coul}$ for
protons:

\begin{equation}
H^{WS}=T+V^{WS}(\vec{r};\beta)+
V^{WS}_{so}(\vec{r};\beta)+\frac{1}{2}(1+\tau_3)V_{Coul}(\vec{r};\beta)\,.
 \end{equation}

\noindent
In the above equation

\begin{equation}
V^{WS}(\vec{r};\beta) =\frac{V_0[1 \pm \kappa(N-Z)/(N+Z)]}
{1+\exp[dist(\vec{r};\beta)/a]}
\end{equation}
and
\begin{equation}
V^{WS}_{so}(\vec{r};\beta)=
-\lambda (\nabla V^{WS}\times\vec{p})\cdot\vec{s}\,,
\end{equation}

\noindent
where $dist(\vec{r};\beta)$ denotes the distance of a point $\vec{r}$ from
the surface of the nucleus given by Eq. (2) and $V_0$, $\kappa$, $a$, $\lambda$
are adjustable constants. The Coulomb potential $V_{Coul}$ is assumed to be 
that of the nuclear charge equal to $(Z-1)e$ and uniformly distributed inside
the nuclear surface. In our calculations we used Woods--Saxon
Hamiltonian with the so--called ``universal'' set of its parameters 
(see Ref. \cite{CDN}) which were
adjusted to the single--particle levels of odd--A nuclei 
with A$\geq$40.

The term $\delta{}E_{\rm pair}$ in Eq.(8) arises from the pairing residual
interaction, which is included to our $H_{s.p.}$ by the BCS approximation.
In the presented paper we used the pairing strength constants:
$G_ZA=13.3+0.217(N-Z)$ and $G_NA=19.3-0.080(N-Z)$ for protons and neutrons,
respectively, which are taken from Ref. \cite{DMS}.

\subsection{Lifetimes for alpha--decay} 
For calculation of alpha--decay half--life we employ the phenomenological
formula of Viola and Seaborg \cite{VS}

\begin{equation}
\log T^{\alpha}_{1/2}\,[yr] =
(aZ + b) (Q_{\alpha}/MeV)^{-1/2} + (cZ +d ) - 7.5\,,
\end{equation}

\noindent
where $Z$ is the atomic number of the parent nucleus and
Q$_{\alpha}$ is the energy release obtained from the mass excesses 

\begin{equation}
Q_{\alpha}(Z,N)=M(Z,N)-M(Z-2,N-2)-M(2,2)\,.
\end{equation}

\noindent
The values of parameters: $a$=1.66175, $b$=--8.5166, $c$=--0.20228
and $d$=--33.9069 in the above formula were taken from Ref. \cite{SPC}.
It should be noted that,
the uncertainties in the calculated  $\alpha$-decay
half-lives due to their phenomenological character are far less than
uncertainties in the calculated SF half-lives. 

\subsection{Lifetimes for spontaneous--fission} 
The spontaneous--fission half--life is inversely proportional to the 
probability of penetration through the barrier

\begin{equation}
    T^{SF}_{1/2}=\frac{\ln2}{n}\frac{1}{P}\,.
    \label{tsf}
\end{equation}

\noindent
Where $n$, in the above formula, is the number of ``assaults''
of the nucleus on the fission barrier {\em per} unit time. 
The number of assaults is usually
equated to the frequency of zero--point vibration of the nucleus in the 
fission degree of freedom and for a vibrational frequency of 
$\hbar\omega_0$=1MeV, assumed in this paper, 
$n\approx10^{20.38}{\rm s}^{-1}$.
Using the one--dimensional WKB semi--classical approximation for 
the penetration probability $P$ one obtains

\begin{equation}
 T^{SF}_{1/2}\,[yr] = \frac{10^{-28.04}}{\hbar\omega_0}\,[1+\exp2S(L)]\,,
\end{equation}

\noindent
where $S(L)$ is the action--integral calculated along a fission path
$L(s)$ in the multi--dimensional deformation space $\{X_\lambda\}$

\begin{equation}
 S(L) = \int^{s_2}_{s_1} \left\{{2 \over \hbar^2} \, B_{\rm eff}(s)
 [V(s) - E]\right\}^{1/2} ds\,.
\label{act}
\end{equation}

\noindent
An effective inertia associated with the fission motion
along the path $L(s)$ is

\begin{equation}
  B_{\rm eff}(s) = \sum_{k,l} \, B_{kl} \,
  {dX_k \over ds} {dX_l \over ds}\,,
\label{beff}
\end{equation}

\noindent
where $B_{kl}$ are the components of the inertia tensor.

In the above equations $ds$ denotes an element of the path length in
the $\{X_\lambda\}$ space. The integration limits $s_1$ and
$s_2$ correspond to the classical turning points,
determined by a~condition $V(s) = E$, where 
$E = V(X^0_{\lambda})$ + 0.5 $\hbar\omega_0$
denotes the energy of the fissioning nucleus in MeV (calculated in the
ground--state).

\subsection{Calculations technique}
Dynamic calculations of $T^{SF}_{1/2}$ mean a quest for 
the least--action trajectory $L_{\rm min}$ which fulfills a principle 
of the least--action $\delta[S(L)]~=~0$.
To minimize the action--integral (\ref{act}) we used the
dynamic--programming method. Its application to fission was first
developed by Baran {\em et al.} (see e.g., \cite{BP} and references
cited therein).

In contrast to the method used by Smola\'nczuk {\em et al.} in Ref.
\cite{SK,SS}, where only two 
coordinates ($\beta_2$ and $\beta_4$) have been handled dynamically 
and the remaining degrees of freedom have been found by
minimization of the potential energy V, in our
multi--dimensional 
dynamic--programming (MDP) method all coordinates are treated
as independent dynamical variables.

Figure 1 demonstrates how our model works.
Since the macroscopic--microscopic method is not analytical, it is
necessary to calculate the potential energy and all components of 
the inertia tensor on a~grid in the multi--dimensional space spanned
by a set of deformation parameters $\{X_\lambda\}$.
We select one coordinate $X_0$ from this set. This
coordinate (e.g. elongation parameter) is related in a linear
way to the fission process.
In Fig.~1 to each point $X_0$ correspond a {\em plane} representing
the rest of the collective space $\{X_{\lambda-1}\}$. 

\begin{figure}[h]
\vspace*{-3.5cm}
\hspace{-4.0cm}
\psfig{file=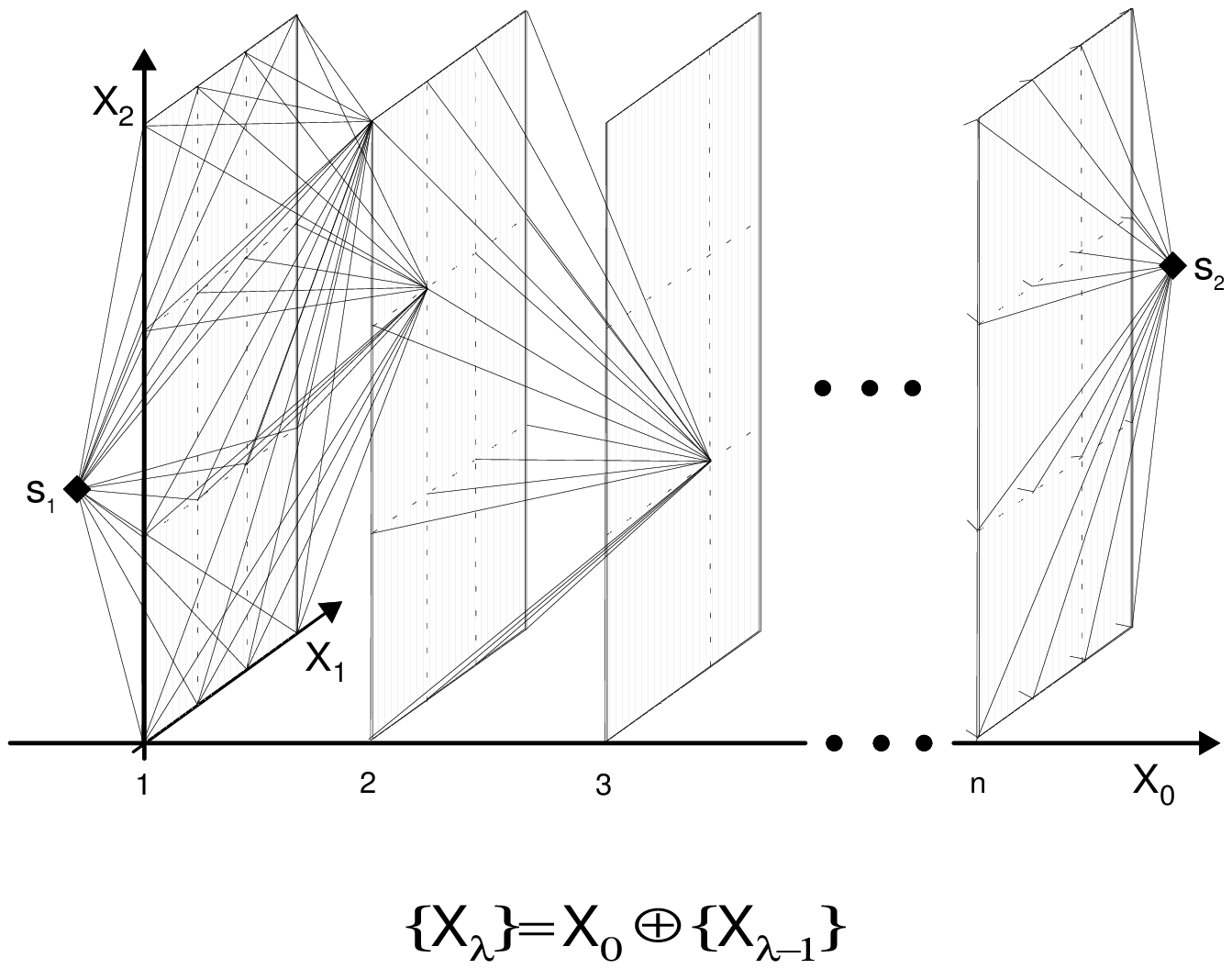,height=24cm}
\vspace{-11.5cm}
\caption{
Diagrammatic illustration of the MDP method.
In the multi--dimensional deformation
space $\{X_\lambda\}$ we select the coordinate $\{X_0\}$
which is related in a linear way to the fission process. The points $s_1$
and $s_2$ correspond to entrance to the barrier and exit from the
barrier, respectively. See text for details.
}
\label{Fig.1}
\end{figure}

To find the least--action trajectory $L_{\rm min}$ between the
turning point $s_1$ and $s_2$ we proceed as follows. First,
we calculate the action--integrals
from the entrance point under the barrier $s_1$
to all grid points in the nearest {\em plane}
at $X_0 = 1$. In the next step we come to the {\em plane} at $X_0 = 2$
and from each grid point in this {\em plane} calculate the
action--integrals to all grid points in the {\em plane} at $X_0 = 1$.
The trajectories started from each grid point at
$X_0 = 2$, passing through all grid points in the {\em plane} at $X_0 = 1$
and terminated in the point $s_1$, form a~bunch of paths. From
each such a~bunch we choose the path with the minimal
action--integral and bear it in mind. At the end of this
step we have the least--action integrals along trajectories
which connect the starting point $s_1$ with all grid points in
the {\em plane} at $X_0 = 2$. Next, we repeat this procedure
for all grid points at $X_0 = 3$ and again we obtain all
the least--action--integrals along trajectories starting from point
$s_1$ with ends at each grid point in the {\em plane} $X_0 = 3$. We
repeat it until we reach the $n$--th {\em plane}, the last one before 
the exit point from the barrier $s_2$. Finally, we proceed to the
last step of our method, where we calculate action--integrals between
the exit point $s_2$ and all grid points 
situated on the last {\em plane} at $X_0 = n$;
the minimal one among them corresponds to the searched
trajectory of the least--action--integral $L_{\rm min}$.

If we denote a number of grid points on each $X_i$ (i=1,2,...,$\lambda$-1)
axis by $n_i$, then the whole number of trajectories examined in MDP
method is equal to 
($n_1\cdot{}n_2\cdot...\cdot{}n_{\lambda-1}$)$^{\textstyle n}$.
Up to now, our calculations are carried out in a maximum of four--dimensional
deformation space in view of an enormously large computational time 
(and disk space) required for preparing (and storing) input data
with potential energy and $1/2\lambda(\lambda+1)$ components of symmetric
inertia tensor for each of 
($n_1\cdot{}n_2\cdot...\cdot{}n_{\lambda-1}$)$\cdot{}n$ grid points.

Calculations are performed in various four--, three-- and two--dimensional
deformation spaces spanned by selected shape parameters
($\beta_{\lambda}$, with
$\lambda$=2,3,4,5,6 and 8) and two pairing degrees of freedom
($\Delta_p, \Delta_n$).
For $\beta$--shape parameters we used grids with steps 
$\Delta \beta_2$=$\Delta \beta_3$=0.05 and $\Delta \beta_{\lambda}$=0.04
for $\lambda$=4,5,6 and 8; in the case of pairing
energy--gaps grids steps are equal 0.2 MeV.
In our calculations a~quadrupole deformation $\beta_2$
plays a~role of the coordinate $X_0$ in Fig.~1.

\section{RESULTS AND DISCUSSION}

\subsection{Optimal Multi--dimensional Deformation Space}
The experimental values of the spontaneous fission
half--lives of nine even--even Fm isotopes (N = 142,
144, ..., 158) form approximately two sides of an acute--angled
triangle with a vertex in N = 152.
This strong nonlinear behaviour of $T^{SF}_{1/2}$ {\em vs.} neutron
number N, due to an enhanced nuclear stability in the vicinity of
deformed shell N=152, 
provides good opportunity for testing theoretical models.

To find the proper deformation space for description of the fission
process we examined three different effects:
the effect of the higher even--multipolarity shape
parameters $\beta_6$ and $\beta_8$,
the role of the reflection--asymmetry shape parameters
$\beta_3$ and $\beta_5$, and 
the influence of the pairing degrees of freedom
$\Delta_p$ and $\Delta_n$.

The following conclusions can be drawn from our previous dynamical
analysis of $T^{SF}_{1/2}$ for Fm even--even isotopes,
see Ref. \cite{LS1,SL1,LS2}.
In the case when the $\beta_{6}$ deformation parameter is 
added to our minimal two--dimensional space $\{\beta_{2}, \beta_{4}\}$ 
we observed an increase of fission lifetimes by one to four orders 
of magnitude. 
The contribution of parameter
$\beta_{8}$ to $T^{SF}_{1/2}$ is negligible.

The deformations with odd--multipolarity $\lambda$=3,5 do not change
SF half--lives.
The reason of this lies in the dynamical treatment of the fission process.
The parameters $\beta_3$ and $\beta_5$ reduce the width of a static
fission barrier, however the effective inertia 
$B_{eff}$, Eq. (\ref{beff}), along the corresponding static path
is larger than along the dynamic one, where $\beta_{3}$ and $\beta_{5}$
are almost equal to zero. One can say, that the static path (corresponding
to minimal potential energy) is ``longer'' than the dynamical one, for which
$\beta_{3}$=$\beta_{5}$=0. The above conclusions are in agreement with
those published in Ref. \cite{SK}.

The pairing degrees of freedom $\Delta_p$ and $\Delta_n$
reduce SF half--lives for Fm isotopes  with N$>$152 for about 3 orders 
of magnitude and considerably improve theoretical predictions of
$T^{SF}_{1/2}$. This effect is due to strong dependence of the inertia
tensor upon pairing energy--gap, as it was shown in Eq. (\ref{delb}).

Finally, we can conclude that the optimal deformation space for 
description of the fission half--lives of heavy nuclei
is 
$\beta_{2}, \beta_{4}, \beta_{6}, \Delta_{p}$ and $\Delta_{n}$.

\begin{figure}[h]
\vspace{-4.5cm}
\centerline{\psfig{file=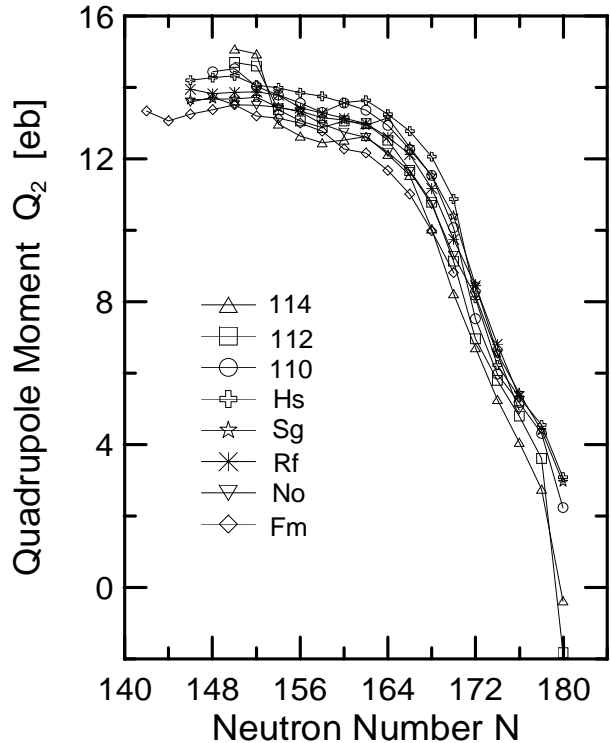,height=20cm}}
\vspace{-5.5cm}
\caption{
The electric quadrupole moments of the even--even nuclei with atomic numbers
Z=100--114, plotted as a function of the neutron number.
}
\label{Fig.2}
\end{figure}

On account of computational limitations mentioned above, we can only perform 
calculations in a maximum of four--dimensional deformation space.
So, we decided to define a correction to SF half--lives, which arises
from pairing degrees of freedom, as the difference between $T^{SF}_{1/2}$
calculated in four--dimensional space
$\{\beta_{2}, \beta_{4}, \Delta_{p}, \Delta_{n}\}$
(when pairing degrees of freedom are treated as dynamical variables)
and the one calculated in two--dimensional space $\{\beta_{2}, \beta_{4}\}$
(where pairing energy--gaps are treated in the stationary way--
i.e. by solving the BCS equations):

\begin{equation}
 \delta T^{SF}_{1/2} (\Delta_{p},\Delta_{n}) \equiv
 T^{SF}_{1/2} (\beta_{2},\beta_{4},\Delta_{p},\Delta_{n}) -
 T^{SF}_{1/2} (\beta_{2},\beta_{4})\,.
\label{tdelta}
\end{equation}
\noindent

Calculations of $T^{SF}_{1/2}$ in the 
space $\{\beta_{2}, \beta_{4}, \beta_{6}, \Delta_{p}, \Delta_{n}\}$
were approximated by the results obtained in 
three--dimensional space $\{\beta_{2}, \beta_{4}, \beta_{6}\}$
with the pairing correction $\delta T^{SF}_{1/2} (\Delta_{p},\Delta_{n})$.

\subsection{Ground--state properties}

\begin{figure}[hb]
\vspace{-3.5cm}
\centerline{\psfig{file=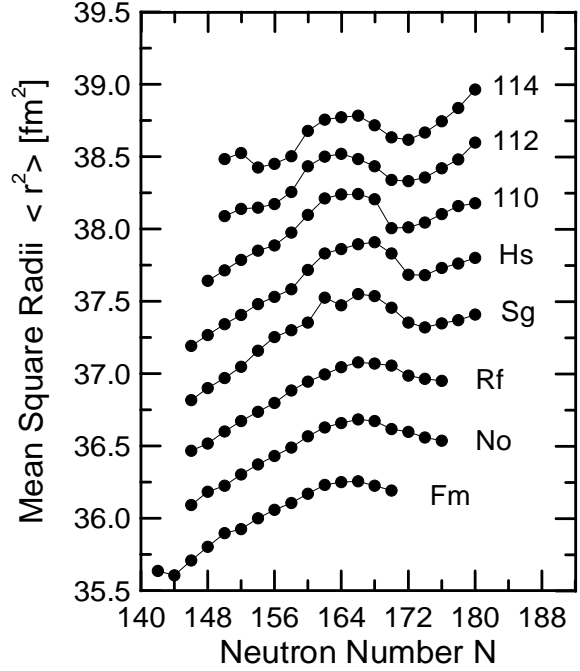,height=18cm}}
\vspace{-5.0cm}
\caption{
The mean square radii of the even--even superheavy nuclei, plotted
as a function of the neutron number.
}
\label{Fig.3}
\end{figure}

In the present study of superheavy the even--even nuclei with
atomic numbers Z=100-114 and neutron numbers N=142-180 are considered.
First, we present the results related to ground--state (GS) properties.
The GS properties were calculated in the equilibrium point found for
a~given nucleus by minimization of its potential energy with respect to
$\beta_{2}$, $\beta_{4}$ and $\beta_{6}$ degrees of freedom.

In Fig.~2 we plot the electric quadrupole moments calculated
with following formula

\begin{equation}
 Q_2 = \sqrt{\frac{16\pi}{5}}
 \sum_{\nu=p} \langle\,\nu\,|\,r^2 Y_{20}\,|\,\nu\,\rangle\, v^2_{\nu}\,,
\end{equation}

\noindent
where $v^2_{\nu}$ is the BCS occupation factor corresponding to 
proton single--particle state $| \nu \rangle$ in the equilibrium point.

Almost all nuclei have distinct prolate deformations.
And with an increase in the neutron number the $Q_2$ values show
a regular decrease except at N=162--164, where a slight 
discontinuity in this behaviour can be seen for nuclei
with atomic number Z$\geq$106.

The mean square charge radii (MSR), for the same region of nuclei,
are plotted as a function of neutron number in Fig.~3.
For calculations of MSR we use the usual formula

\begin{equation}
 <r^2> = \frac{1}{\rm Z}
 \sum_{\nu=p} \langle\,\nu\,|\,r^2\,|\,\nu\,\rangle\, v^2_{\nu}
 + 0.64\,{\rm fm^2}\,,
\end{equation}

\noindent
where the last term is due to finite range of proton charge distribution.

One observes a rather regular dependence of mean square radii 
on both neutron and proton number. However, as previously,
close to N=162--164 one can see local maxima in MSR curves,
particularly for nuclei with Z$\geq$106. 
This means that Coulomb repulsion energy for these nuclei is
locally smaller, then they are more stable.

\subsection{Spontaneous--fission versus alpha--decay}

\begin{figure}[h]
\vspace{-6.0cm}
\centerline{\psfig{file=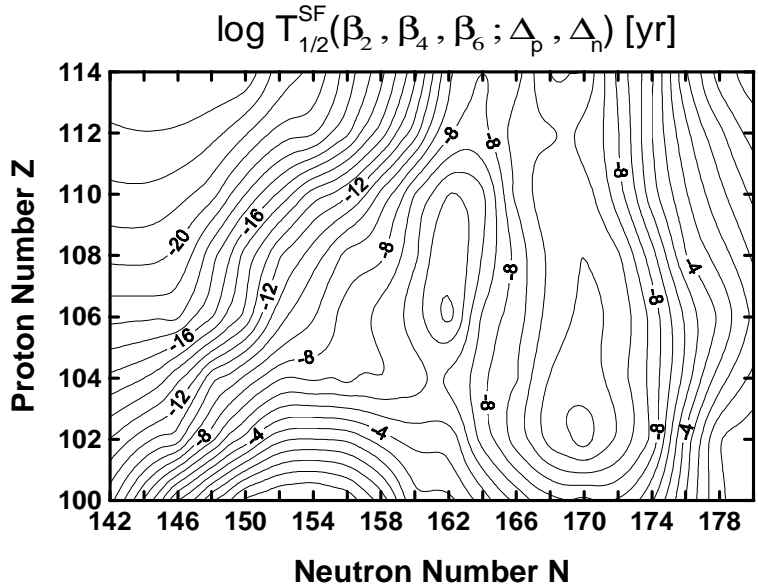,height=20cm}}
\vspace{-6.0cm}
\caption{
Contour map of the logarithm of the spontaneous--fission
half-life (given in years) for nuclei shown in Fig.~2 
calculated in $\{\beta_2, \beta_4, \beta_6\}$ deformation space
and corrected by the effect of the pairing degrees of freedom,
Eq. (18).
}
\label{Fig.4}
\end{figure}

Figure 4 shows the results of the spontaneous--fission
half--lives calculation, according to the MDP method, performed in 
$\{\beta_{2}, \beta_{4}, \beta_{6}\}$ deformation space
with pairing correction $\delta T^{SF}_{1/2} (\Delta_{p},\Delta_{n})$,
Eq. (\ref{tdelta}), for the nuclei shown in Fig.~2.

Two very specific effects can be observed on the contour map of
$T^{SF}_{1/2}$ plotted as a function of neutron and proton numbers.
One can see an enhancement in the SF half--life values at N=162
followed by a diminution at N=170.

The enhancement in nuclear stability near the deformed shell N=162
allows the appearance of a peninsula of deformed metastable 
superheavy nuclei. The local maximum of the $T^{SF}_{1/2}$ values
is centered at the nucleus $^{268}_{106}{\rm Sg}_{162}$ (2.5 h).

In the vicinity of neutron number N=170 one observes the opposite behaviour.
Here, the SF half--life values form a trench. This trench separates the 
peninsula of the deformed superheavy nuclei from an island of spherical
superheavy elements around the doubly magic nucleus
$^{298}114_{184}$. The local minimum is obtained for nucleus
$^{272}_{102}{\rm No}_{170}$ (10 $\mu$s).

We found also that the $T^{SF}_{1/2}$ values of two heaviest nuclei
(considered in the presented paper)
$^{292}114_{178}$ and $^{294}114_{180}$ are comparable with those
of the most stable Fm isotopes with neutron numbers N=152 and~154
($\sim$ 100 yr).

In Fig.~5 we plot the alpha--decay half--lives (given in years)
estimated by use of the Viola--Seaborg relationship with set 
of constants from Ref. \cite{SPC}. The contour plot of the
$T^{\alpha}_{1/2}$ forms a relatively regular surface descending
steeply in the direction where the proton number tends to increase 
and the neutron number tends to decrease (upper--left--hand
corner in the plot).

\begin{figure}[h]
\vspace{-6.0cm}
\centerline{\psfig{file=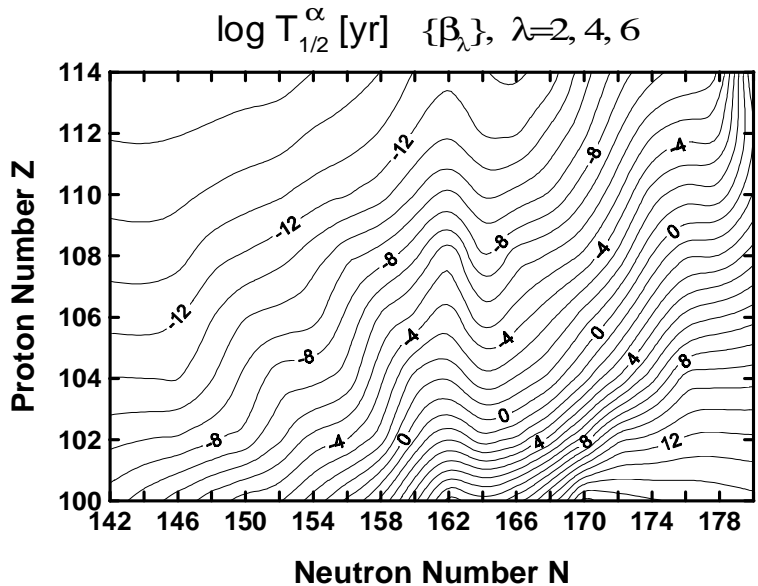,height=20cm}}
\vspace{-6.0cm}
\caption{
Contour plot of the logarithm of the alpha--decay half-life
$T^{\alpha}_{1/2}$ (given in years) 
obtained from the Viola--Seaborg systematics. The alpha--decay energy
was calculated with $\{\beta_{\lambda}\}, \lambda=2,4,6$
deformation parameters.
}
\label{Fig.5}
\end{figure}

The surface of the $T^{\alpha}_{1/2}$ values shows an evident
protuberance at N=162,
which demonstrates unambiguously the magicity of this neutron 
number. The shell effect at N=152 is very weak and disappears 
practically for nuclei with atomic number Z$>$104.

The results presented in Fig.~4 and~5 as well as  conclusions 
drawn from them are generally similar to those recently
published by other groups employing the 
macroscopic--microscopic method, Ref. \cite{SS,SSS,MN,MNK}.

To compare the SF and $\alpha$-decay modes we show in Fig.~6 the logarithm
of the total half--life $T^{SF+\alpha}_{1/2}$ resulting from both
these modes. If we examine contour maps with $T^{SF+\alpha}_{1/2}$
and $T^{SF}_{1/2}$ we can notice some minor differences but the global
behaviour of both these quantity stays unchanged.

The dark shadowed areas in Fig.~6 show the regions of nuclei in
which the $\alpha$--decay mode predominates.
The light shadowed area corresponds to the intermediate region of
nuclei, where the probabilities of the SF and $\alpha$--decay
processes are approximately equal (i.e. the region where the values of
$T^{\alpha}_{1/2}$ and $T^{SF}_{1/2}$ differ up to one order of 
magnitude).
The black solid curve inside this area connects nuclei for which
probabilities of both considered modes are the same.

Thus, one can observe that the region of increased $\alpha$--decay 
activity separates two areas with predominating SF activity
in a diagonal manner, in Fig.~6.
It is worth noting that the upper area is almost inaccessible due to
extremely short lifetimes ($\leq$ 1 $\mu$s).

\begin{figure}[h]
\vspace{-6.0cm}
\centerline{\psfig{file=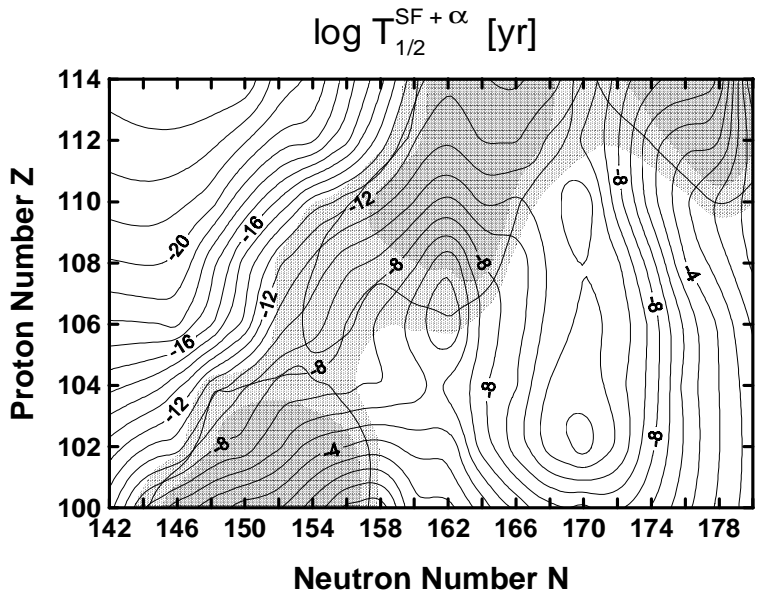,height=20cm}}
\vspace{-6.0cm}
\caption{
Logarithm of the total half-life $T^{SF+\alpha}_{1/2}$ 
resulting from SF (Fig.~4) and $\alpha$--decay (Fig.~5) modes.
The dark shadowed areas show the regions of nuclei, where the $\alpha$--decay
mode is a dominant one. The light shadowed area corresponds to the
intermediate region, where the probabilities of SF and $\alpha$--decay modes
are approximately equal.
}
\label{Fig.6}
\end{figure}

The total half--life values (equal to $T^{\alpha}_{1/2}$) for two
heaviest nuclei, considered in this paper, 
$^{292}114_{178}$ and $^{294}114_{180}$ are found larger than 1 yr.
This is in agreement with results obtained recently in the fully
selfconsistent microscopic nonrelativistic 
Hartrre--Fock--Bogoliubov approach, Ref. \cite{BBD}.


\begin{thebibliography}{99}
\itemsep=0cm

\bibitem{Hof1}
S.~Hofmann et al.,Z. Phys. {\bf A350} (1995) 277.
  
\bibitem{Hof2}
S.~Hofmann et al.,Z. Phys. {\bf A350} (1995) 281.
  
\bibitem{Hof3}
S.~Hofmann et al.,Z. Phys. {\bf A354} (1996) 229.
  
\bibitem{Laz1}
Yu.A. Lazarev et al., Phys. Rev. Lett. {\bf 73} (1994) 624.

\bibitem{Laz2}
Yu.A. Lazarev et al., Phys. Rev. Lett. {\bf 75} (1995) 1903.

\bibitem{T}
A. T\"urler, Proc. Inter. Workshop XXIV, Hirschegg 1996,
ed. H.~Feldmeier, J. Knoll and W.~N\"orenberg, GSI Darmstadt 1996, p. 29

\bibitem{LS1}
Z.~\L{}ojewski and A.~Staszczak, Acta. Phys. Pol. {\bf B27} (1996) 531.

\bibitem{SL1}
A.~Staszczak and Z.~\L{}ojewski, Proc. Inter. Workshop XXIV, Hirschegg
1996, ed. H.~Feldmeier, J. Knoll and W.~N\"orenberg, GSI Darmstadt 1996,
p. 98 (Report nucl--th/9603032).
  
\bibitem{LS2}
Z.~\L{}ojewski and A.~Staszczak, Annales Univ. MCS, Sectio AAA,
{\bf L/LI} (1995/1996) 137.

\bibitem{MB}
L.G.~Moretto and R.B.~Babinet, Phys.~Lett.~{\bf 49B} (1974) 147.

\bibitem{SB}
A.~Staszczak, A.~Baran, K.~Pomorski and K.~B\"oning,
Phys. Lett.~{\bf 161B} (1985) 227.

\bibitem{SPP}
A.~Staszczak, S.~Pi\l{}at and K.~Pomorski, Nucl.~Phys.~{\bf A504}
(1989) 589.
 
\bibitem{PPS}
S.~Pi\l{}at and K.~Pomorski and A.~Staszczak, Proceed. Confer. Fifty Years
with Nuclear Fission, vol.II, ed. J.W. Behrens, A.D. Carlson, American Nuclear
Society, Gaithersburg 1989, p.637

\bibitem{D}
J.~Dudek, Report CRN/PHTH 91-19, Strasbourg 1991.

%\bibitem{Ing}
%D.R.~Inglis, Phys. Rev. {\bf 96} (1954) 1059; {\bf 97} (1955) 701.

\bibitem{Belya}
S.T.~Belyaev, Mat. Fys. Medd. Dan. Vid. Selsk. {\bf 30} No 11 (1959);
Nucl. Phys. {\bf 24} (1961) 322.

\bibitem{GP}
A.~G\'o\'zd\'z, K.~Pomorski, M.~Brack and E.~Werner, Nucl. Phys. {\bf A442}
(1985) 26.

\bibitem{Sob69}
A.~Sobiczewski, Z.~Szyma\'nski, S.~Wycech, S.G.~Nilsson,
J.R.~Nix, C.F.~Tsang, C.~Gustafson, P.~M\"oller and
B.~Nilsson, Nucl. Phys. {\bf A131} (1969) 67. 

\bibitem{Funny}
M.~Brack, J.~Damgaard, A.S.~Jensen, H.C.~Pauli, V.M.~Strutinsky
and C.Y.~Wong, Rev. Mod. Phys. {\bf 44} (1972) 320.

\bibitem{Strut}
V.~M.~Strutinsky, Nucl. Phys. {\bf A95} (1967) 420;
Nucl. Phys. {\bf A122} (1968) 1.

\bibitem{KN}
H.J.~Krappe, J.R.~Nix and A.J.~Sierk, Phys.~Rev.~{\bf C20} (1979) 992.

\bibitem{CDN}
S.~\'Cwiok, J.~Dudek, W.~Nazarewicz, J.~Skalski and T.~Werner,
Comput. Phys. Commun.~{\bf 46} (1987) 379.
 
\bibitem{DMS}
J.~Dudek, A.~Majhofer and J.~Skalski, J. Phys. {\bf G6} (1980) 447.

\bibitem{VS}
V.E.~Viola, Jr. and G.T.~Seaborg, J. Inorg. Nucl. Chem. {\bf 28} (1966) 741.

\bibitem{SPC}
A.~Sobiczewski, Z.~Patyk and S.~\'Cwiok, Phys. Lett. {\bf B224} (1989) 1. 

\bibitem{BP}
A.~Baran, K.~Pomorski, A.~\L{}ukasiak and A.~Sobiczewski, 
Nucl. Phys. {\bf A361} (1981) 83. 

\bibitem{SK}
R.~Smola\'nczuk, H.V.~Klapdor--Kleingrothaus and A.~Sobiczewski,
Acta Phys. Pol. {\bf B24} (1993) 685.

\bibitem{SS}
R.~Smola\'nczuk, J.~Skalski and A.~Sobiczewski, Phys. Rev. {\bf C52}
(1995) 1871.

\bibitem{SSS}
R.~Smola\'nczuk, J.~Skalski and A.~Sobiczewski, Proc. Inter. Workshop XXIV,
Hirschegg 1996, ed. H.~Feldmeier, J. Knoll and W.~N\"orenberg, GSI Darmstadt
1996, p. 35.

\bibitem{MN}
P.~M\"oller and J.R.~Nix, J. Phys. {\bf G20} (1994) 1681

\bibitem{MNK}
P.~M\"oller, J.R.~Nix and K.L. Kratz, At. Data. Nucl. Data Tables,
to be published (Report nucl-th/9601043).

\bibitem{BBD}
J.-F.~Berger, L.~Bitaud, J.~Decharg\'e, M.~Girod and S.~Peru--Desenfants,
Proc. Inter. Workshop XXIV, Hirschegg 1996, ed. H.~Feldmeier, J. Knoll 
and W.~N\"orenberg, GSI Darmstadt 1996, p. 43.

\end{thebibliography}
\end{document}